\documentclass[12pt]{article}
\usepackage{a4wide}
\usepackage{graphicx}

\newcommand{\be}{\begin{equation}}
\newcommand{\ee}{\end{equation}}
\newcommand{\ba}{\begin{eqnarray}}
\newcommand{\ea}{\end{eqnarray}}
\newcommand{\mrm}[1]{\mathrm{#1}}

\newcommand{\Q}{{\cal Q}}

\begin{document}

\begin{titlepage}
\begin{flushright}
LU TP 04-37\\
hep-ph/0410333\\
revised November 2004\\
\end{flushright}
\vfill
\begin{center}
{\Large\bf Isospin Breaking in $K\to3\pi$ Decays II: \\ 
\vspace{0.5 cm}
Radiative Corrections}

\vfill

{\bf Johan Bijnens and Fredrik Borg}\\[1cm]
{Department of Theoretical Physics, Lund University\\
S\"olvegatan 14A, S 22362 Lund, Sweden}
\end{center}

\vfill

\begin{abstract}
The five different CP conserving amplitudes for the
decays $K\to3\pi$ are calculated using Chiral Perturbation 
Theory. The calculation is made to next-to-leading order and includes 
full isospin breaking. The squared amplitudes are compared
with the corresponding ones in the isospin limit to estimate the size of the 
isospin breaking effects. In this paper we add the radiative corrections
to the earlier calculated $m_u-m_d$ and local electromagnetic effects. 
We find corrections of order 5-10 percent.  
\end{abstract}
\vfill

\end{titlepage}

\section{Introduction}
\label{introduction}

The non-perturbative nature of low-energy QCD calls for alternative methods
of calculating processes including composite particles such as 
mesons and baryons. 
A method describing the interactions of the light pseudoscalar mesons 
($K,\pi,\eta$) is Chiral Perturbation Theory (ChPT). It was introduced by 
Weinberg, Gasser and Leutwyler \cite{Weinberg,GL1,GL2} and 
it has been very successful. Pedagogical introductions 
to ChPT can be found in \cite{chptlectures}. 
The theory was later
extended to also cover the weak interactions of the pseudoscalars \cite{KMW1}, 
and the first calculation of a kaon decaying into pions ($K\to2\pi,3\pi$) 
appeared shortly 
thereafter \cite{KMW2}. Reviews of other applications of ChPT to 
nonleptonic weak interactions can be found in \cite{chptweakreviews}. 
 
A recalculation in the isospin limit of $K\to2\pi$ to next-to-leading order
was made in \cite{BPP,BDP} and
of $K\to3\pi$ in \cite{BDP,GPS}.
In \cite{BDP} also a full fit to all 
experimental data was made and it was found that the decay rates and 
linear slopes agreed well. However, a small discrepancy was
found in the quadratic slopes and that is part of the motivation for this
further investigation of the decay $K\to3\pi$ in ChPT.

The discrepancies found can have several different origins. It could be an 
experimental problem or it could have a theoretical origin. In the latter 
case the corrections to the amplitude calculated in \cite{BDP} are threefold:
strong isospin breaking, electromagnetic (EM) isospin breaking or higher order
corrections. These effects have been studied in many papers for the
$K\to2\pi$ decays, references can be traced back from \cite{K2PIiso}.
For $K\to3\pi$ less work has been done.
In \cite{BB} the strong isospin and local electromagnetic corrections were
investigated and it was found that the inclusion of those led to changes of
a few percent in the amplitudes. The local electromagnetic part was
also calculated in \cite{GPS}, in full agreement with our result 
after corrections of some misprints in \cite{GPS}.

In this paper we add also the radiative
corrections, i.e.\ the nonlocal electromagnetic isospin breaking. 
The full (first order) isospin 
breaking amplitude to 
next-to-leading chiral order is thus calculated,
and we will try to estimate
the effect of this in the amplitudes. A new full fit, including also 
new experimental data \cite{istra,kloe}, has to be done to 
answer the question whether isospin breaking 
removes the problem of fitting the quadratic slopes. This, together with
a study of models for the higher order coefficients, we plan to do in an 
upcoming paper. 

Other recent results on
$K\to3\pi$ decays can be found in \cite{cabibbo,nehme}. In \cite{cabibbo}
Nicola Cabibbo discusses the possibility of determining the $a_0-a_2$ pion 
scattering length from the threshold effects of $K^+\to\pi^0\pi^0\pi^+$. He
gives an approximate theoretical result with very few unknown parameters. 
We have a possibly better theoretical description of these effects but it
includes more unknown parameters. In 
\cite{nehme} an attempt was made to calculate the virtual photon
corrections to the $K^+\to\pi^0\pi^0\pi^+$ decay. Our result 
disagrees with the result presented there.

The outline of this paper is as follows. The next section describes isospin 
breaking in more detail. In section \ref{lagrangian} the basis of ChPT, 
the Chiral Lagrangians, are discussed. Section \ref{kinematics} 
specifies the decays and describes the relevant kinematics. The
divergences appearing when including photons are discussed in section 
\ref{isoIR}. In section \ref{isoanalyticres} the analytical results are 
discussed, section \ref{isonumres} contains the numerical results and the
last section contains the conclusions. 

\section{Isospin Breaking}
\label{striso}

Isospin symmetry is the symmetry under exchange of up-
and down-quarks. Obviously this symmetry is only true in the approximation
that $m_u=m_d$ and electromagnetism is neglected, i.e.\ in the isospin limit.
Calculations are often performed in the isospin limit since this is 
simpler and gives a good first estimate of the result.

However, to get a precise result one has to include isospin 
breaking, i.e.\ the effects from $m_u\neq m_d$ and electromagnetism. 
Effects coming 
from $m_u\neq m_d$ we refer to as strong isospin breaking and 
include mixing between $\pi^0$ and
$\eta$. This mixing leads to changes in the formulas for both the 
physical masses of
$\pi^0$ and $\eta$ as well as the amplitude for any process involving either 
of the two. For a detailed discussion see \cite{strongiso}.

The other source is electromagnetic isospin breaking, coming from the
fact that the up- and the down-quarks are charged, which implies 
different interactions with photons. This part can be further divided in local 
electromagnetic isospin breaking and explicit photon contributions
(radiative corrections). 
The former are described
by adding new Lagrangians at each order and the latter by introducing 
new diagrams including photons.  

Our first calculation of $K\to3\pi$ \cite{BDP} was done in the isospin limit.
In the next paper, \cite{BB}, we included strong and local isospin breaking
(there collectively referred 
to as strong isospin breaking) and we now present the calculation including 
all isospin breaking effects.

\section{The ChPT Lagrangians}
\label{lagrangian}

The basis of our ChPT calculation is the various  Chiral Lagrangians. 
They can be divided in different orders. The 
order parameters in the perturbation series are $p$ and $m$, the momenta and
mass of the pseudoscalars. Including isospin breaking also $e$, the electron
charge, and the mass difference, $m_u-m_d$, are used as order parameters.
All of these are independent expansion parameters. We work to
leading order in $m_u-m_d$ and $e^2$ but next-to-leading order in
$p^2$ and $m^2$. For simplicity we call 
in the remainder terms of order 
$p^2$, $m^2$, $e^2$ and $m_u-m_d$ leading order, and terms 
of order $p^4$ ,$p^2\,m^2$, $m^4$,
$p^2\,e^2$, $m^2\,e^2$, $p^2 (m_u-m_d)$ and $m^2 (m_u-m_d)$
next-to-leading order.
 
\subsection{Leading Order}
\label{leading}

The leading order Chiral Lagrangian is usually divided in three parts
\be
{\cal L}_{2}={\cal L}_{S2}+{\cal L}_{W2}+{\cal L}_{E2},
\ee
where ${\cal L}_{S2}$ refers to the strong $\Delta S = 0$ part, 
${\cal L}_{W2}$ the weak 
$\Delta S = \pm 1$ part, and ${\cal L}_{E2}$ the strong-electromagnetic and 
weak-electromagnetic parts combined. For the strong part we have \cite{GL1}
\be
\label{L2S}
{\cal L}_{S2} = \frac{F_0^2}{4} \; \langle u_\mu u^\mu + \chi_+\rangle
\ee
Here $\langle A\rangle$ stands for the flavour trace of the matrix $A$,
and $F_0$ is the pion decay constant in the chiral limit.
We define the matrices $u_\mu$, $u$ and $\chi_\pm$ as
\be
u_\mu = i u^\dagger\, D_\mu U\, u^\dagger = u_\mu^\dagger\,,\quad u^2 = U\,,
\quad \chi_\pm = u^\dagger \chi u^\dagger \pm u \chi^\dagger u\,,
\ee
where the special unitary matrix $U$ contains the Goldstone boson fields
\be
U = \exp\left(\frac{i\sqrt{2}}{F_0}M\right)\,,\quad
M =\left(\begin{array}{ccc}
\frac{1}{\sqrt{2}}\pi_3+\frac{1}{\sqrt{6}}\eta_8 & \pi^+ & K^+\\
\pi^- & \frac{-1}{\sqrt{2}}\pi_3+\frac{1}{\sqrt{6}}\eta_8 & K^0\\
K^- & \overline{K^0} & \frac{-2}{\sqrt{6}}\eta_8
         \end{array}\right)\,.
\ee
The formalism we use is the external field method of \cite{GL1}, 
and to include photons we set
\be
\chi = 2 B_0
\left(\begin{array}{ccc}m_u &  & \\ & m_d & \\ & & m_s\end{array}\right)\, 
\,\,\,\, \mrm{and}\,\,\,\, D_\mu U = \partial_\mu U - i e\, Q A_\mu U
 - i e\, U Q A_\mu,
\ee
where $A_\mu$ is the photon field and 
\begin{equation} \label{defZ}
Q = \left( \begin{array}{ccc} 2/3 & &\\ & -1/3 & \\ & & -1/3
\end{array}
\right).
\ee
We diagonalize the quadratic terms in (\ref{L2S}) by a rotation
\ba
\pi^0 &=& \pi_3\cos\epsilon + \eta_8\sin\epsilon\,
\nonumber\\
\eta &=& -\pi_3\sin\epsilon + \eta_8\cos\epsilon\,,
\ea
where the lowest order mixing angle $\epsilon$ satisfies
\be
\tan(2\epsilon) = \sqrt{3}\frac{m_d-m_u}{2\,m_s-m_u-m_d}\,.
\ee

The weak part of the Lagrangian has the form \cite{Cronin}
\be
\label{LW2}
{\cal L}_{W2} = C \, F_0^4 \, 
\Bigg[ G_8 \langle \Delta_{32} u_\mu u^\mu \rangle +
G_8' \langle\Delta_{32} \chi_+ \rangle  +
G_{27} t^{ij,kl} \, \langle \Delta_{ij } u_\mu \rangle
\langle\Delta_{kl} u^\mu \rangle \Bigg]
+ \mbox{ h.c.}\,.\nonumber
\ee
The tensor $t^{ij,kl}$ has as nonzero components
\ba
\label{deft}
t^{21,13} =
t^{13,21} = \frac{1}{3} \, &;& \, 
t^{22,23}=t^{23,22}=-\frac{1}{6} \, ; \nonumber \\
t^{23,33}=t^{33,23}=-\frac{1}{6} \, &;& \, 
t^{23,11} =t^{11,23}=\frac{1}{3}\,,
\ea
and the matrix $\Delta_{ij}$ is defined as
\be
\Delta_{ij} \equiv u \lambda_{ij} u^\dagger\,,\quad
\left(\lambda_{ij}\right)_{ab} \equiv \delta_{ia} \, \delta_{jb}\,.
\ee
The coefficient $C$ is defined such that in the chiral
and large $N_c$ limits $G_8 = G_{27} =1$,
\be
C= -\frac{3}{5} \, \frac{G_F}{\sqrt 2} V_{ud} \, V_{us}^* = 
-1.06\cdot 10^{-6}\,\, \mrm{GeV}^{-2}\, .
\ee
Finally, the remaining electromagnetic part, relevant for this calculation,
 looks like (see e.g. \cite{EMEcker})
\be
\label{LE2}
{\cal L}_{E2} = 
e^2 F_0^4 Z \langle {\cal Q}_L {\cal Q}_R\rangle +
e^2 F_0^4 \langle \Upsilon {\cal Q}_R\rangle
\ee
where the weak-electromagnetic term is characterized by a 
constant $G_E$ \mbox{($g_{\rm ewk}G_8$ in \cite{EMEcker})},
\begin{equation} \label{gewk}
\Upsilon = G_E\, F_0^2 \Delta_{32} + {\rm h.c.}  \, 
\ee
and 
\begin{equation}
\Q_L = uQu^\dagger\,,\,\,\, \Q_R = u^\dagger Qu \, .
\ee

\subsection{Next-to-leading Order}
\label{ntleading}

The fact that ChPT is a non-renormalizable theory means that new terms 
have to be added at 
each order to compensate for the loop-divergences. This means that
the Lagrangians increase in size for every new order and the number of 
free parameters rises as well.
At next-to-leading order the Lagrangian is split in four parts which, in
obvious notation, are 
\be
{\cal L}_{4}={\cal L}_{S4}+{\cal L}_{W4}+{\cal L}_{S2E2}+{\cal L}_{W2E2}(G_8)
\,.
\ee
Here the notation $(G_8)$ indicates that here only the dominant
$G_8$-part is included in the Lagrangian and therefore in the calculation.

These Lagrangians are quite large and we choose not to write them explicitly 
here since they can be found in many places 
\cite{GL1, radiative,KMW1,Esposito,EKW,EMEcker,urech}. 
For a list of all the pieces relevant for this specific calculation 
see \cite{BB}.
Note however that four terms producing photon interactions should be added to
${\cal L}_{W4}$ in \cite{BB}. The two new terms in the octet part are
\be
N_{14}\, i\, \langle\Delta_{32}\, \{f_+^{\mu\nu},u_\mu u_\nu\}  \rangle \, +
N_{15}\, i\, \langle\Delta_{32}\, u_\mu f_+^{\mu\nu} u_\nu  \rangle
\ee
and in the 27 part
\be
D_{13}\, i\, t^{ij,kl} \langle\Delta_{ij}\, u_\mu \rangle 
                       \langle\Delta_{kl}\, [u_\nu,f_+^{\mu\nu}]\rangle \, +
D_{15}\, i\, t^{ij,kl} \langle\Delta_{ij}\, u_\mu u_\nu \rangle
                       \langle\Delta_{kl}\, f_+^{\mu\nu} \rangle,
\ee  
where
\be
f_+^{\mu\nu} = u F^{\mu\nu} u^\dagger + u^\dagger F^{\mu\nu} u,\,\,\, 
F^{\mu\nu} = e\, Q\, (\partial^\mu A^\nu - \partial^\nu A^\mu).
\ee  

\subsubsection{Ultraviolet Divergences}

The process $K\to 3\pi$ receives 
higher-order contributions from diagrams that contain loops. 
The study of these diagrams is complicated
by the fact that they need to be defined precisely. The loop-diagrams involve
an integration over the undetermined loop-momentum $q$, and the 
integrals are divergent in the $q\rightarrow\infty$ ultraviolet region. 
These ultraviolet divergences are canceled by
replacing the coefficients in the next-to-leading order Lagrangians by
the renormalized coefficients and a subtraction part, see \cite{BDP,BB}
and references therein.
The divergences can be used as a check on the calculation and all our
infinities (except the ones left since the $G_{27}$-part in
${\cal L}_{W2E2}$ is not known) 
cancel as they should.

\subsubsection{Loop Integrals}

The prescription we use for the loop integrals can be found in
many places, e.g. \cite{ABT1}. The only one needed in addition to the ones
given there is the one-loop three point function
\be
C(m_1^2,m_2^2,m_3^3,p_1^2,p_2^2,p_3^2) =
\frac{1}{i}\int \frac{d^d p}{(2\pi)^d} \frac{1}{\left(p^2-m_1^2\right)
\left((p-p_1)^2-m_2^2\right)\left((p-p_3)^2-m_3^2\right)}\,,
\ee
where $p_3=p_1+p_2$.
For its numerical evaluation we use the program FF \cite{FF}.
This program also deals with possible infrared divergences consistently.

\section{Kinematics}
\label{kinematics}

There are five different CP-conserving decays of the type $K\to3\pi$
($K^-$ decays are not treated separately since they are counterparts
to the $K^+$ decays):
\ba
\label{defdecays}
K_L(k)&\to&\pi^0(p_1)\,\pi^0(p_2)\,\pi^0(p_3)\,,\quad [A^L_{000}]\,,\nonumber\\
K_L(k)&\to&\pi^+(p_1)\,\pi^-(p_2)\,\pi^0(p_3)\,,\quad [A^L_{+-0}]\,,\nonumber\\
K_S(k)&\to&\pi^+(p_1)\,\pi^-(p_2)\,\pi^0(p_3)\,,\quad [A^S_{+-0}]\,,\nonumber\\
K^+(k)&\to&\pi^0(p_1)\,\pi^0(p_2)\,\pi^+(p_3)\,,\quad [A_{00+}]\,,\nonumber\\
K^+(k)&\to&\pi^+(p_1)\,\pi^+(p_2)\,\pi^-(p_3)\,,\quad [A_{++-}]\,,
\ea
where we have indicated the four-momentum defined for each particle
and the symbol used for the amplitude.

The kinematics is treated using
\be
s_1 = \left(k-p_1\right)^2\,,\quad
s_2 = \left(k-p_2\right)^2\,,\quad
s_3 = \left(k-p_3\right)^2\,.
\ee
The amplitudes are expanded in terms of the Dalitz plot variables
$x$ and $y$ defined as
\be
\label{defsi}
y =\frac{ s_3-s_0}{m_{\pi^+}^2}\,,\quad
x =\frac{ s_2-s_1}{m_{\pi^+}^2}\,,\quad
s_0 = \frac{1}{3}\left(s_1+s_2+s_3\right)\,.
\ee
The amplitude for  $K_L\to\pi^0\pi^0\pi^0$ is
symmetric
under the interchange of all three final state particles and the one for
$K_S\to\pi^+\pi^-\pi^0$ is antisymmetric under the interchange of $\pi^+$ and
$\pi^-$ because of CP.
The amplitudes for
 $K_L\to\pi^+\pi^-\pi^0,K^+\to\pi^+\pi^+\pi^-$ and $K^+\to\pi^0\pi^0\pi^+$
are symmetric under the interchange of the first two pions because of
CP or Bose-symmetry. 

\section{Infrared Divergences}
\label{isoIR}

In addition to the ultraviolet divergences which are removed by
renormalization, diagrams including photons in the loops contain infrared (IR)
divergences. These
infinities come from the $q\rightarrow 0$ end of the loop-momentum 
integrals. They are canceled by including also the 
Bremsstrahlung diagram, where a real photon
is radiated off one of the charged mesons, see Fig.~\ref{brems}.
It is only the sum of the virtual loop corrections and the real Bremsstrahlung
which is physically significant and thus needs to be well defined.
\begin{figure}
\begin{center}
\includegraphics[width=0.4\textwidth]{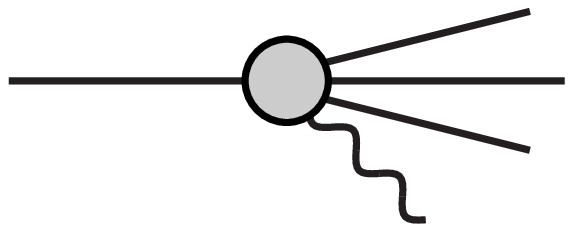}
\end{center}
\caption{Bremsstrahlung, the emission of an extra final-state photon.
\label{brems}}
\end{figure}
We regulate the IR divergence in both the virtual photon loops and the real
emission with a photon mass $m_\gamma$ and keep only the singular terms
plus those that do not vanish in the limit $m_\gamma\to 0$. We include the 
real Bremsstrahlung for photon energies up to a cut-off $\omega$ and
treat it in the soft photon approximation.

The exact
form of the amplitude squared for the bremsstrahlung diagram 
depends on which specific amplitude that is 
being calculated. For $K^+(k) \to \pi^0(p_1) \pi^0(p_2) \pi^+(p_3)$ 
it can be written in the soft photon limit
(see e.g. \cite{Peskin})
\be
|A|^2_{BS} = |A|_{LO}^2 \, e^2 \, \int \frac{d^3q}{(2\pi)^3}\,\frac{1}{2q}\, 
\sum_{\lambda=0,1}
\left[\frac{k\cdot \epsilon^{(\lambda)}}{q\cdot k} - 
      \frac{p_3\cdot \epsilon^{(\lambda)}}{q\cdot p_3}\right ]^2,  
\ee
where $|A|_{LO}$ is the lowest order isospin limit amplitude.
The number of terms inside the parentheses is the number of charged
particles in the process and the sign of those terms
depends both on the charge of the radiating particle and on whether it is 
incoming or outgoing. Writing out the square and using 
$\sum_{\lambda=0,1} \epsilon^{(\lambda)}_\mu \epsilon^{(\lambda)}_\nu 
= -g_{\mu \nu}$, you get 
\be
|A|^2_{BS} = -|A|_{LO}^2 \, e^2 \, \int \frac{d^3q}{(2\pi)^3}\,\frac{1}{2q}\, 
\left [\frac{k^2}{(q\cdot k)^2} + \frac{p_3^2}{(q\cdot p_3)^2} 
- \frac{2\, p_3\cdot k}{(q\cdot k)(q\cdot p_3)} 
\right ]
\ee
To solve the first integral term, place the vector $k$ along the z-axis, i.e.\
\be
k = (k^0,0,0,k^z)\,\,\, \mrm{and}\,\,\,(k\cdot q)^2 = (k^0q^0-k^zq^z)^2. 
\ee
Changing to polar coordinates that part of the integral now looks like
\be
-|A|_{LO}^2 \, e^2 \, \frac{m_K^2}{8\pi^2} \int dq\, 
d(\cos\theta)
\, \frac{q}{(k^0 E_\gamma-k^z q \cos\theta)^2},
\ee
where $k^2=m_K^2$, $q_0=E_\gamma$ and $q_z=q \cos\theta$ have been used.
Solving the $d(\cos\theta)$ part is now straightforward and leads to
\be
-|A|_{LO}^2 \, e^2 \, \frac{m_K^2}{8\pi^2} \int dq
\, \frac{1}{k^z}\left(\frac{1}{k^0 E_\gamma-k^z q}-
\frac{1}{k^0 E_\gamma+k^z q}\right ),
\ee
Putting the two terms on a common denominator and changing variable to
$E_\gamma$ leads to
\be
-|A|_{LO}^2 \, e^2 \, \frac{m_K^2}{4\pi^2} \int^\omega_{m_\gamma} 
dE_\gamma
\frac{E_\gamma}{E_\gamma^2 (k^0)^2-(E_\gamma^2-m_\gamma^2) (k^z)^2}\,,
\ee
where $\omega$ is the photon energy above which the detector identifies
it as a real external photon.  
We are only interested in the result in the limit $m_\gamma\to 0$, so 
it's enough to consider
\be
-|A|_{LO}^2 \, e^2 \, \frac{m_K^2}{4\pi^2} \int^\omega_{m_\gamma}
 dE_\gamma
\frac{1}{m_K^2 E_\gamma},
\ee
which gives the result
\be
-|A|_{LO}^2 \, \frac{e^2}{8\pi^2} \log \frac{\omega^2}
{m_\gamma^2}\,.
\ee
In a similar way one gets the result for the mixed term
\be
\int \frac{d^3q}{(2\pi)^3}\,\frac{1}{2q}\, 
\left [ 2 \frac{p_3\cdot k}{(q\cdot k)(q\cdot p_3)} 
\right ] = -\frac{x_s}{4\pi^2} \frac{s_3-m_K^2-m_\pi^2}
{m_K m_\pi (1-x_s^2)}
\log x_s \log \frac{\omega^2}
{m_\gamma^2} \equiv I_{IR}(m_K^2,m_\pi^2,s_3)\,, 
\ee
where
\be
x_s = \frac{\sqrt{1-4 m_K m_\pi / (\bar{s}_3-(m_K-m_\pi)^2)}-1}
{\sqrt{1-4 m_K m_\pi / (\bar{s}_3-(m_K-m_\pi)^2)}+1}.
\ee
In order to
obtain the correct imaginary part we use the $i\varepsilon$-prescription,
which means $\bar{s}_3 = s_3+i\varepsilon$.

For the other amplitudes the calculations are similar and 
the resulting bremsstrahlung amplitudes are
\be
|A^L_{000}|^2_{BS} = 0 \,,
\ee
\be
|A^L_{+-0}|^2_{BS} = 
-|A_{+-0}^L|^2_{LO} \,\frac{e^2}{4\pi^2}\,\left[  \log 
\frac{\omega^2}{m_\gamma^2} - I_{IR}(m_\pi^2,m_\pi^2,s_3)\right]\,,
\ee
\be
|A^S_{+-0}|^2_{BS} = 
-|A^S_{+-0}|^2_{LO} \,\frac{e^2}{4\pi^2}\,\left[  \log 
\frac{\omega^2}{m_\gamma^2} -I_{IR}(m_\pi^2,m_\pi^2,s_3)\right]\,,
\ee
\be
|A_{00+}|^2_{BS} = 
-|A_{00+}|^2_{LO} \,\frac{e^2}{4\pi^2}\,\left[  \log 
\frac{\omega^2}{m_\gamma^2}  -I_{IR}(m_\pi^2,m_K^2,s_3) \right]\,,
\ee
\ba
|A_{++-}|^2_{BS} & =  &  -|A_{++-}|^2_{LO} \,\frac{e^2}{4\pi^2}\,
\left[  2 \log \frac{\omega^2}{m_\gamma^2}
 -I_{IR}(m_\pi^2,m_K^2,s_1) -I_{IR}(m_\pi^2,m_K^2,s_2)
 \right.\nonumber \\&&
 +I_{IR}(m_\pi^2,m_K^2,s_3)
 -I_{IR}(m_\pi^2,m_\pi^2,s_1) -I_{IR}(m_\pi^2,m_\pi^2,s_2)
\nonumber\\&&
\left.
 +I_{IR}(m_\pi^2,m_\pi^2,s_3)\right]\,.
\ea
When using the above, the divergences
from the explicit photon loops cancel exactly. 

A similar problem shows up in the definition of the decay constants since
we normalize the lowest order with $F_{\pi^+}$ and $F_{K^+}$.
Our prescription for the decay constants is described in App.~\ref{App:fpi}.

\section{Analytical Results}
\label{isoanalyticres}

\subsection{Lowest order}

The four diagrams that could contribute to lowest order can be seen in
Fig.~\ref{isofigtree}.
\begin{figure}
\begin{center}
\includegraphics[width=0.99\textwidth]{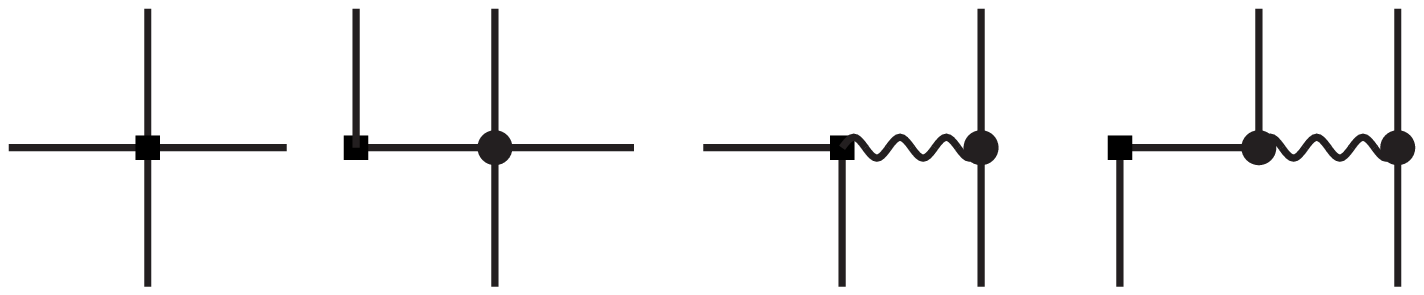}
\end{center}
\caption{The tree level diagrams for $K\to3\pi$. A filled square is a 
weak vertex, a filled circle a strong vertex, a straight line a pseudoscalar
meson and a wiggly line a photon.
\label{isofigtree}}
\end{figure}

However, the two diagrams including photons turn out to give zero. 
This is obviously so for $K^+\to\pi^+\pi^0\pi^0$ and $K_L\to\pi^0\pi^0\pi^0$
since the $\gamma\pi^0\pi^0$ vertex vanishes as a consequence of charge
conjugation.

The reason why it vanishes for the other decays is somewhat more subtle
and is the same as why the lowest order result for $K\to\pi\ell^+\ell^-$
vanishes \cite{EPR}.
When doing a simultaneous diagonalization
of the covariant kinetic and mass terms quadratic in the pseudoscalar fields, 
including those of the weak lagrangian ${\cal L}_{W2}$, $p^2$-terms of the
form $\partial_\mu K \partial^\mu \pi$ are absent and all weak vertices 
involve at least three pseudoscalar fields. This result should
not change as compared to our calculation where the weak Lagrangian was not
included in the diagonalization.   
Thus in our case, the two diagrams on the right in Fig.~\ref{isofigtree} will 
together give zero contribution.

This means that the lowest order result in the full isospin case is the same
as when just including strong and local EM isospin breaking. This result
we published before, the full expressions can be found in \cite{BB}.  

\subsection{Next-to-leading order}

There are 51 additional diagrams contributing to
next-to-leading order. They can be divided in three different classes and 
examples 
will be shown of each class. It should be noted that the argument in the
previous subsection is not valid at this order.
There now exist $K\pi\gamma$ vertices.
The reason for this is that one can not diagonalize simultaneously all
terms with two pseudoscalar fields when going to next-to-leading order.

The first class of diagrams are the 13 which do not include explicit photons. 
They are the ones used in our earlier papers \cite{BDP,BB} and a complete
list of them can be found there. Some examples are shown in 
Fig.~\ref{figp4}.
\begin{figure}
\begin{center}
\includegraphics[width=0.99\textwidth]{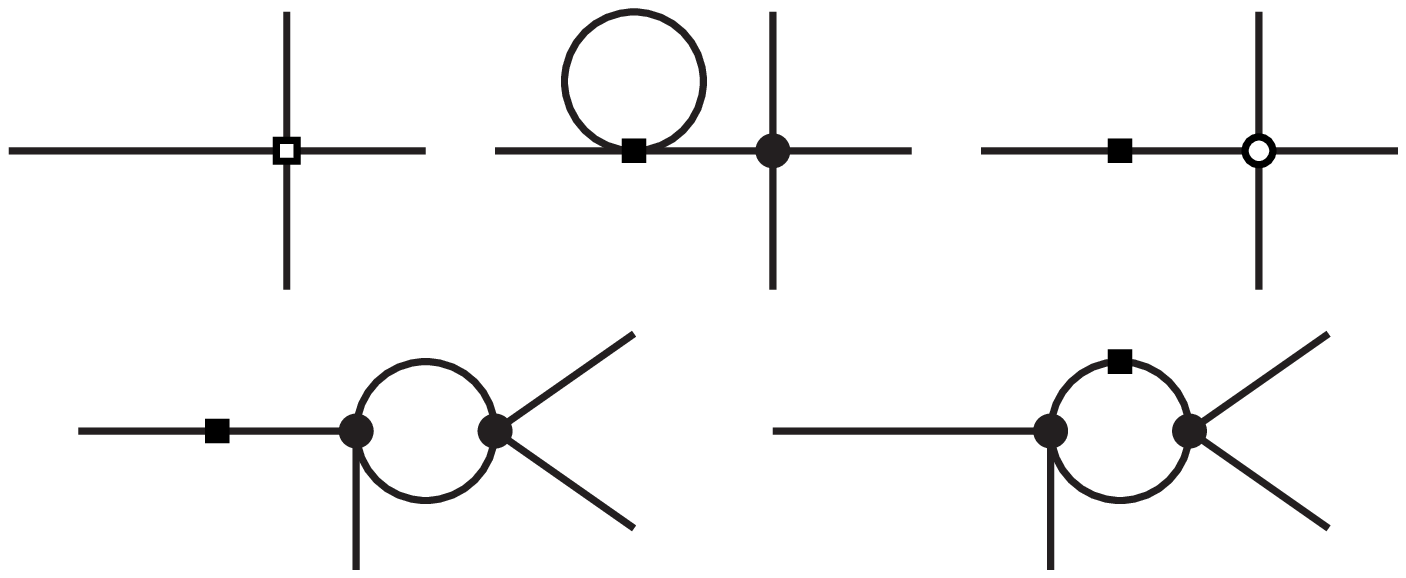}
\caption{Examples of diagrams of next-to-leading order with no photons.
An open square is a vertex
from ${\cal L}_{W4}$ or ${\cal L}_{W2E2}$, an open circle a vertex from 
${\cal L}_{S4}$ or ${\cal L}_{S2E2}$,
a filled square a vertex
from ${\cal L}_{W2}$ or ${\cal L}_{E2}\, (\Delta S = 1)$ and a filled 
circle a vertex 
from ${\cal L}_{S2}$ or ${\cal L}_{E2}\, (\Delta S = 0)$.
\label{figp4}}
\end{center}
\end{figure}

The second class of diagrams are the ones with a photon running in a
loop. There are 18 of these and some examples can be found in
Fig.~\ref{loopfigp4}. Their evaluation is the main new result of
this paper. They are also responsible for the infrared divergences
discussed in Sect.~\ref{isoIR}. 
\begin{figure}
\includegraphics[width=0.9\textwidth]{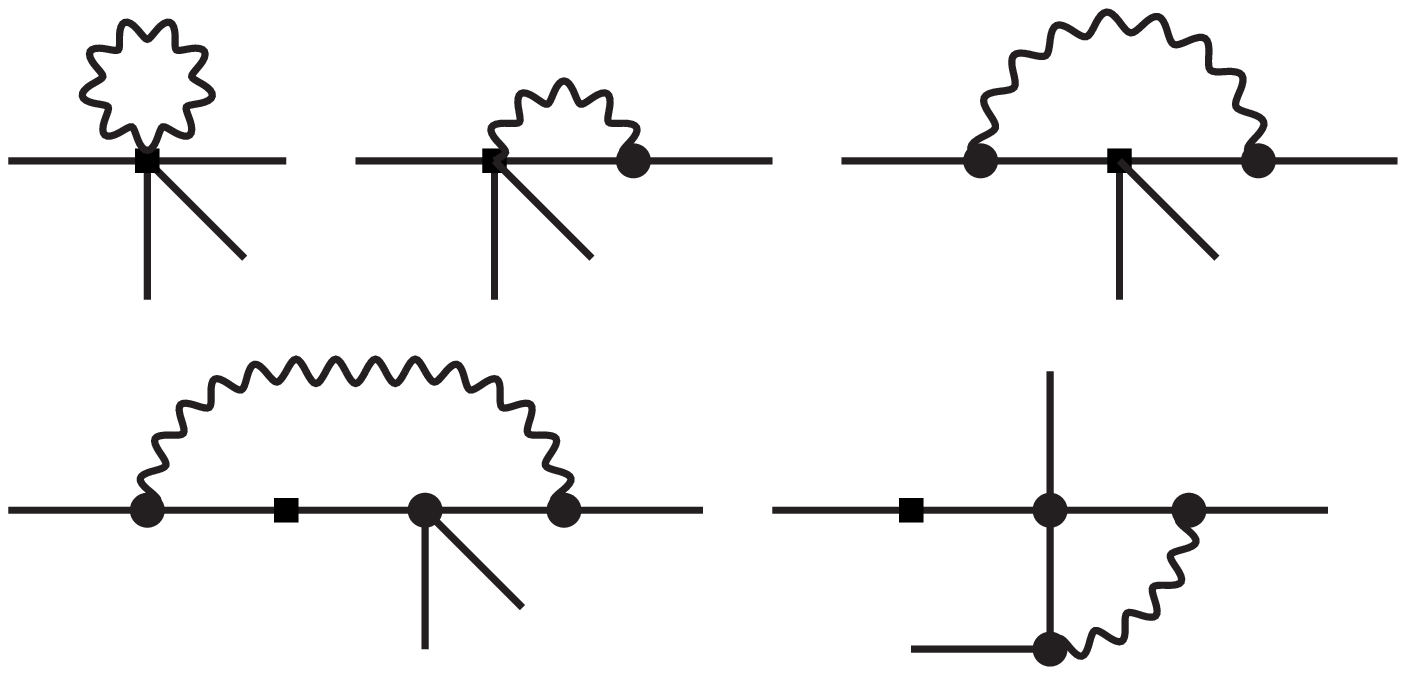}
\begin{center}
\end{center}
 \caption{Examples of diagrams with photons in the loops. A filled square is a 
weak vertex, a filled circle a strong vertex, a straight line a pseudoscalar
meson and a wiggly line a photon.
\label{loopfigp4}}
\end{figure}
The first diagram is an example where the photon is the only particle in
the loop, a photon tadpole diagram. These vanish in dimensional regularization
when only singular and nonzero terms in the limit $m_\gamma\to 0$ are kept.

The last class of diagrams is the ones with tree level photon propagators, 20
in total. They are photon reducible, i.e.\ if we cut the photon line the 
diagram falls apart.
They are 
infrared finite and some examples can be seen in Fig.~\ref{propfigp4}.
It turns out that for realistic values of the input parameters this class
of diagrams give a negligible contribution to all $K\to3\pi$ processes.
\begin{figure}
\begin{center}
\includegraphics[width=0.95\textwidth]{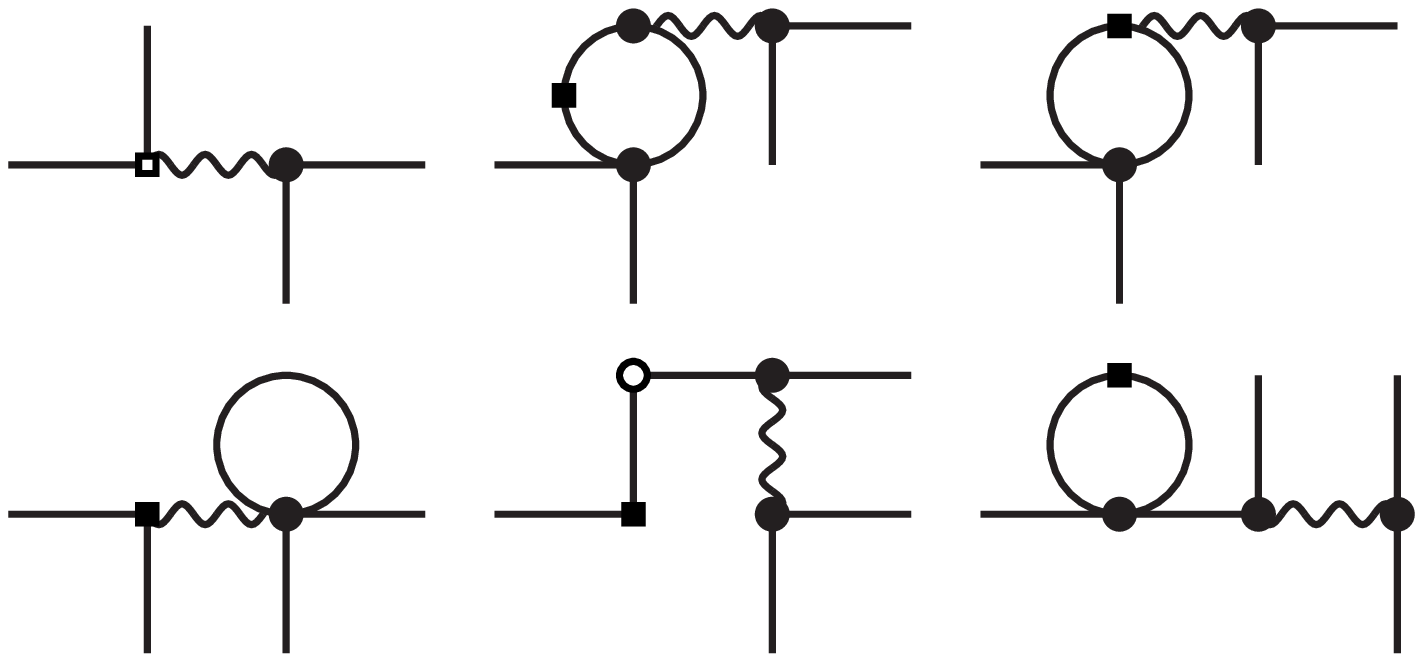}
\end{center}
\caption{Examples of diagrams with photon propagators. An open square 
is a vertex
from ${\cal L}_{W4}$ or ${\cal L}_{W2E2}$, an open circle a vertex from 
${\cal L}_{S4}$ or ${\cal L}_{S2E2}$,
a filled square a vertex
from ${\cal L}_{W2}$ or ${\cal L}_{E2}\, (\Delta S = 1)$ and a filled 
circle a vertex 
from ${\cal L}_{S2}$ or ${\cal L}_{E2}\, (\Delta S = 0)$. A  straight line is 
a pseudoscalar meson and a wiggly line a photon.
\label{propfigp4}}
\end{figure}

We work to first order in isospin breaking, i.e.\ as soon as $e^2$ is present
we set $m_u=m_d$, $m_{\pi^+} = m_{\pi^0} = m_\pi$ and 
$m_{K^+} = m_{K^0} = m_K$. Even then,
the resulting amplitudes at next-to-leading order are rather long,
and it does not seem very useful to present them here
explicitly. However, the full 
expressions for the amplitudes are available on request from the 
authors or can be downloaded \cite{formulas}.
It should be noted that the 
amplitude for $K_L\to \pi^0\pi^0\pi^0$ was given in \cite{BB}. It was stated
that it was the full isospin breaking amplitude since 
no explicit photon diagrams can contribute to this process. 
This is not completely true,
the amplitude will change indirectly through the definition 
of $F_{\pi^+}$ and $F_{K^+}$, see App.~\ref{App:fpi}. 

The next-to-leading order amplitudes can in principle 
include all the low-energy coefficients (LECs) from  
${\cal L}_{W4}$, ${\cal L}_{S4}$, 
${\cal L}_{W2E2}$ and ${\cal L}_{S2E2}$. The coefficients from the strong
and electromagnetic part of the Lagrangian are treated as input, which leaves
$G_8$,$G_{27}$,$N^r_1,\ldots,N^r_{13}$, $D^r_1$,$D^r_2$,$D^r_4,\ldots,D^r_7$,
$D^r_{26},\ldots,D^r_{31}$ and $Z^r_1,\ldots,Z^r_{14}$ as 
undetermined. 
In total 41 unknown parameters. However, all of these do not appear
independently, i.e.\ they multiply the same type of term, e.g. 
$m_K^4$ or $e^2 m_{\pi}^2$. It turns out, as discussed in \cite{BB},
 that there are 30 independent 
combinations, denoted by $\tilde K_{1} \ldots \tilde K_{30}$. For the 
11 combinations already appearing in the isospin limit see \cite{BDP,BB}
and the 19 additional ones for the isospin breaking case can be
be found in \cite{BB}.
In addition to the 30 combinations found in \cite{BB},
four new coefficients show up when including photons: $N^r_{14}$, $N^r_{15}$,
$D^r_{13}$ and $D^r_{15}$. These all come from the third class of diagrams
where a tree level photon is present.
The four coupling constants show up in precisely the same combinations
in $K\to\pi\ell^+\ell^-$ and can thus be determined experimentally
in other decays. We therefore treat them as input.

\section{Numerical Results}
\label{isonumres}

\subsection{Experimental data and fit}

A full isospin limit fit was made in \cite{BDP} taking into account all
data published before May 2002. One of the motivations for this 
continued investigation 
of isospin breaking effects is to see whether isospin violation can solve 
the discrepancies in the quadratic slope parameters found there. A new full 
fit will be done in an upcoming paper. 
The data from ISTRA+ \cite{istra} and KLOE \cite{kloe}, which 
appeared after \cite{BDP}, will then also be taken into account.
We do not present a new fit in this paper since estimates of the
new combinations of constants should be done before attempting a full fit.

\subsection{Inputs}

\begin{table}
\begin{center}
\begin{tabular}{||c|r||c|r||c|r||}
\hline
\hline
$G_8$   & $5.45$ &$L_1^r$ & $0.38\cdot10^{-3}$ & $\tilde K_1 $ & $0$ \\
$G_{27}$ & $0.392$ & $L_2^r$ & $1.59\cdot10^{-3}$ & 
$\tilde K_2/G_8 $ & $5.19\cdot 10^{-2}$ \\
$G_E$   & $-0.4$ & $L_3^r$ & $-2.91\cdot10^{-3}$ & 
$\tilde K_3/G_8 $ & $3.77\cdot 10^{-3}$ \\
                  &        & $L_4^r$ & $0$ & $\tilde K_4 $ & 0 \\
$\sin\epsilon$ & $1.19\cdot 10^{-2}$&$L_5^r$ & $1.46\cdot10^{-3}$ & 
$\tilde K_5/G_{27} $ & $-4.25\cdot 10^{-2}$ \\
$Z$ & $0.805$ & $L_6^r$ & $0$ & 
$\tilde K_6/G_{27} $ & $-1.66\cdot 10^{-1}$ \\
$\mu$& 0.77 GeV&$L_7^r$ & $-0.49\cdot10^{-3}$ & $\tilde K_7/G_{27} $ & 
$1.20\cdot 10^{-1}$ \\
$F_\pi$&$0.0924$ GeV& $L_8^r$ & $1.0\cdot10^{-3}$ &
  $\tilde K_8\ldots \tilde K_{11} $ & 0 \\ 

$F_K$&$0.113$ GeV&$L_9^r$& $7.0\cdot10^{-3}$ & 
$\tilde K_{12}\ldots \tilde K_{30} $ & $0$ \\
$N_{14}$&$-10.4\cdot 10^{-3}$&$K_1\ldots K_{11}$&$0$&$D_{13}$&$0$ \\
$N_{15}$&$5.95\cdot 10^{-3}$&&&$D_{15}$&$0$ \\
\hline
\hline
\end{tabular}
\end{center}
\caption{The various input values used.\label{tab:isoInput}}
\end{table}

The input values we use are presented 
in Table~\ref{tab:isoInput}.

\subsubsection{Strong and Electromagnetic Input}

There are different ways to treat the masses, especially in the isospin limit 
case. In \cite{BDP} the masses used in the phase space were obtained
from the physical masses
occuring in the decays. However in the amplitudes the physical
mass of the kaon involved in the process was used and the pion 
mass was given by $m_{\pi}^2=\frac{1}{3}\sum_{i=1,3} m_{\pi^i}^2$
with $i=1,2,3$ being the three pions participating in the reaction. 
This allowed for the correct kinematical relation 
$s_1+s_2+s_3= m_K^2+3m_\pi^2$ to be satisfied while having the isospin
limit in the amplitude but the physical masses in the phase space.
The results in \cite{BDP} were obtained with the physical mass for the
eta.  Results with the Gell-Mann-Okubo (GMO) relation for the
eta mass in the loops gave small
changes within the general errors given in \cite{BDP}. 

In the decays here, we work to first order in isospin breaking. We have
rewritten explicit factors of $m_u-m_d$ in terms of $\sin\epsilon$ according
to
\be
m_u-m_d = -\frac{2}{\sqrt{3}}\,\, (2\,m_s-m_u-m_d)\,
\sin\epsilon \,.
\ee
In  general we use the physical masses of pions and kaons in the loops but
as soon as
a factor of $\sin\epsilon$ or $e^2$ is present we use a common kaon and
a common pion mass. This simplifies the analytical formulas enormously.
The kaon mass chosen is the mass from the kaon
in the decay and the pion mass used is $3 m_\pi^2 = \sum_i m_{\pi^i}^2$
with $i=1,2,3$ the three pions in the final state, i.e.\ the mass we used
in the isospin limit case. For the eta mass we use in general the GMO mass
in the loops but with isospin violation included,
\be
m_\eta^2 = \frac{2}{3}\, (m_{K^+}^2+m_{K^0}^2-m_{\pi^+}^2) + 
\frac{1}{3}\, m_{\pi^0}^2\,.
\ee
The possible lowest order contributions from the eta mass have been removed
from the amplitudes using the corresponding next-to-leading order relation
as described in \cite{BB}.

The strong LECs
$L_1^r$ to $L_8^r$ from ${\cal L}_{S4}$ as well as $\sin\epsilon$ 
come from the one-loop fit in \cite{strongiso}. 

The constant $Z$ from ${\cal L}_{E2}$ we estimate via
\be
Z = \frac{1}{2\,F_\pi^2\,e^2}\, (m_{\pi^+}^2-m_{\pi^0}^2),
\ee
which corresponds to the value in Table~\ref{tab:isoInput}. 
The higher order coefficients of ${\cal L}_{E4}$, $K_1\ldots K_{11}$,
 are rather unknown.
Some rough estimates exist but we put them to zero here.

The IR divergences are cancelled by adding the soft-photon Bremsstrahlung.
We have used a 10~MeV cut-off in energy for this and used the same
cut-off in the definition of $F_{\pi^+}$ and $F_{K^+}$.

The subtraction scale $\mu$ is chosen to be $0.77$ GeV. 

\subsubsection{Weak Inputs}

The coefficients contributing in the isospin limit from ${\cal L}_{W2}$
and ${\cal L}_{W4}$ are taken from the fit in \cite{BDP}.
The values of $G_8$, $G_{27}$
in Table~\ref{tab:isoInput} are taken from \cite{BDP} as well.
We use as a reasonable estimate for $G_E$ the value presented in \cite{BP}.

No knowledge exists of the values of
$\tilde K_{12}\ldots \tilde K_{30} $,
 so they are set equal to
zero at $\mu=0.77$~GeV.
Tests were also made assigning order of magnitude estimates 
to them. This imparted changes in the isospin breaking corrections
similar in size to those of the loop contributions.

\subsubsection{Input relevant for the Photon Reducible Diagrams}

The two new constants $D_{13}$ and $D_{15}$ are set to zero since no knowledge
exist of their values. Note that in order to contribute at all, they have to 
get values orders of magnitude larger than the expected size.

The two other new constants, $N_{14}$ and $N_{15}$, 
can be determined from \mbox{$K\to\pi l^+l^-$} decays. For these decays, the
branching ratios can be expressed as \cite{EPR}
\ba
\label{Kpll}
\mrm{BR}(K^+\to\pi^+ e^+ e^-)& = & (3.15-21.1w_++36.1w_+^2)\cdot 10^{-8}\, 
|C\,G_8/(9\cdot 10^{-6}\cdot \mrm{GeV}^{-2})|^2  \nonumber\\
\mrm{BR}(K_S\to\pi^0 e^+ e^-)& = & (3.07-18.7w_S+28.4w_S^2)\cdot 10^{-10}\, 
|C\,G_8/(9\cdot 10^{-6}\cdot \mrm{GeV}^{-2})|^2  \nonumber\\
\ea
Using the measured central values \cite{PDG}
\be
\mrm{BR}(K^+\to\pi^+ e^+ e^-) = 2.88\cdot 10^{-7},\,\,\,
\mrm{BR}(K_S\to\pi^0 e^+ e^-) = 3.0\cdot 10^{-9}
\ee
one gets the results
\be
\label{signchoice}
w_+^{(1)} = 1.69,\,\,\, w_+^{(2)} = -1.10\,\,\,\,\, \mrm{and}\,\,\,\,\, 
w_S^{(1)} = 1.93,\,\,\, w_S^{(2)} = -1.28.
\ee

These constants, $w_+$ and $w_S$, can then be written in terms
of both strong and weak low-energy constants \cite{EPR},
\ba
\label{w+ws}
w_+ & = &  \frac{4}{3}\,(4\pi)^2 [N^r_{14}(\mu)-N^r_{15}(\mu)
+3\,L^r_{9}(\mu)]-\frac{1}{3}\ln{\frac{m_K m_\pi}{\mu^2}}\nonumber\\
w_S & = &  \frac{2}{3}\,(4\pi)^2 [2\,N^r_{14}(\mu)+N^r_{15}(\mu)]
-\frac{1}{3}\ln{\frac{m_K^2}{\mu^2}},
\ea
and from the knowledge of $L^r_{9}(\mu)$ \cite{L9}, one 
gets the values listed in Table~\ref{tab:isoInput} for one choice of signs
in eq.~(\ref{signchoice}).

Note that the influence of this class of diagrams is not visible in the figures
nor in the results shown in the tables for any of the possible signs
chosen in eq.~(\ref{signchoice}).
The magnitude of these constants
needs to be increased significantly in order to have a visible impact on
our numerical results.

\subsection{Results with and without isospin breaking}

The results we will present here is a comparison between the squared amplitudes
and the decay rates in the isospin limit and including first order isospin
breaking. 
The full squared amplitudes over the decay region is
a 3-D plot with the two different
cases plotted over phasespace. This is however very difficult to read, and 
instead we will present comparisons along three slices of these 3-D plots
as explained below. We also present the corrections for the Dalitz plot
parameters.

In Table~\ref{tab:centralvalue} we present the values of the amplitudes
squared in the center of the Dalitz plot, i.e.\ for $x=y=0$.
We show the results in the isospin limit from \cite{BDP}, with the
strong and local electromagnetic isospin breaking included \cite{BB}
and with full isospin breaking included.
\begin{table}
\begin{center}
\begin{tabular}{|c|c|c|c|}
\cline{2-4}
\multicolumn{1}{c|}{} & 
\multicolumn{3}{|c|}{Centralvalue} \\
\cline{2-4}
\multicolumn{1}{c|}{} & 
Iso \cite{BDP} & Strong \cite{BB}& Full \\
\hline
$K_L \to \pi^0 \pi^0 \pi^0$ &
  $6.74 \cdot 10^{-12}$ & $6.97 \cdot 10^{-12}$ & $7.04 \cdot 10^{-12}$ \\
$K_L \to \pi^+ \pi^- \pi^0$ &
 $7.46 \cdot 10^{-13}$ &  $7.66 \cdot 10^{-13}$ & $7.88 \cdot 10^{-13}$ \\
$K_S \to \pi^+ \pi^- \pi^0$ &
 $0$ & $0$ & $0$ \\
$K^+ \to \pi^0 \pi^0 \pi^+$ &
 $0.93 \cdot 10^{-13}$ & $1.01 \cdot 10^{-12}$ & $1.03 \cdot 10^{-12}$ \\
$K^+ \to \pi^+ \pi^+ \pi^-$ &
 $3.72 \cdot 10^{-12}$ & $4.00 \cdot 10^{-12}$ & $4.14 \cdot 10^{-12}$ \\
\hline
\end{tabular}
\end{center}
\caption{Comparison of the central values of the amplitudes squared
 in the isospin conserving case (Iso), including strong and
local electromagnetic (Strong) and full (Full) isospin breaking.
\label{tab:centralvalue}}
\end{table}

Similarly, in Table~\ref{tab:decay} we present the integrated decay rates
in the isospin conserving case \cite{BDP}, the one with strong and local
electromagnetic isospin breaking included \cite{BB} and with all isospin 
breaking effects
included. There are here in principle problems with an infinite correction
when a charged two pion system is at rest. The effects of the 
electromagnetic interaction
can then become very large from terms containing logarithms of the pion 
velocity.
This is where the Coulomb interaction dominates and it should then really be
resummed to all orders. In order to avoid this problem we have introduced the
cut-off $E_C$. It means that we only integrate the phase space
over the part where 
\be
\sqrt{s_i} \ge 2 m_\pi + E_C\,.
\ee
Due to the way we have chosen the pion mass, the Coulomb problem only shows up
for the decay $K^+\to\pi^+\pi^+\pi^-$ where the systems of two charged pions
can be at rest or at very low relative velocity at the edges of phase space.
The places where this happens are indicated by the large dots in
Fig.~\ref{fig:phase}. The choice of the pion masses in
$K_{L,S}\to\pi^+\pi^-\pi^0$ is such that the Coulomb threshold is slightly
outside the physical phasespace. It turns out that in this case the part
of the correction that includes the Coulomb singularity is rather small.
We have therefore not included any corrections for it in the results presented.
\begin{table}
\begin{center}
\begin{tabular}{|c|c|c|c|c|}
\cline{2-4}
\multicolumn{1}{c|}{} & 
 \multicolumn{3}{c|}{Decay Rate} \\
\cline{2-5}
\multicolumn{1}{c|}{} & 
 Iso \cite{BDP}& Strong & Full & $E_C$ [MeV] \\
\hline
$K_L \to \pi^0 \pi^0 \pi^0$ &
  $2.65 \cdot 10^{-18}$ & $2.74 \cdot 10^{-18}$ & $2.77 \cdot 10^{-18}$ & \\
$K_L \to \pi^+ \pi^- \pi^0$ &
 $1.63 \cdot 10^{-18}$ & $1.67 \cdot 10^{-18}$ & $1.72 \cdot 10^{-18}$ &\\
$K_S \to \pi^+ \pi^- \pi^0$ &
 $3.1 \cdot 10^{-21}$ & $3.2 \cdot 10^{-21}$ & $3.3 \cdot 10^{-21}$ &\\
$K^+ \to \pi^0 \pi^0 \pi^+$ &
 $9.11 \cdot 10^{-19}$ & $9.84 \cdot 10^{-19}$ & $1.00 \cdot 10^{-18}$ &\\
$K^+ \to \pi^+ \pi^+ \pi^-$ &
 $2.97 \cdot 10^{-18}$ & $3.19 \cdot 10^{-18}$ & --- & 0\\
& $2.95 \cdot 10^{-18}$ & $3.17 \cdot 10^{-18}$ &$3.28 \cdot 10^{-18}$  & 1\\
& $2.91 \cdot 10^{-18}$ & $3.13 \cdot 10^{-18}$ & $3.24 \cdot 10^{-18}$ & 2\\
& $2.72 \cdot 10^{-18}$ & $2.93 \cdot 10^{-18}$ &$3.03 \cdot 10^{-18}$ & 5\\
\hline
\end{tabular}
\end{center}
\caption{Comparison of the
decay rates in the isospin conserving case (Iso), including strong and
local electromagnetic (Strong) and full (Full) isospin breaking.
The Coulomb cut-off used, $E_C$, is explained in the text.
\label{tab:decay}}
\end{table}

In Fig.~\ref{fig:phase} we also show the phase space boundaries for the five 
different decays and the three curves along which  we will show
results for the squared amplitudes with and without isospin 
breaking. The three curves are $x=0$, $y=0$ and $x = \sqrt{3}\,y$. 
In Fig.~\ref{fig:compare1} to Fig.~\ref{fig:compare5} we then plot the 
five different squared amplitudes along these curves as a 
function of $r$, where
$r=\pm\sqrt{y^2+\frac{x^2}{3}}$ and the sign is chosen according to
\be
r = \left\{ \begin{array}{ll} 
y, & x = 0 \\
x/\sqrt{3}, & y = 0 \\
y\,\sqrt{2}, & x= y\,\sqrt{3}\,.
\end{array}
\right.
\ee
Note that for all but $A^S_{+-0}$ the squared
amplitudes are normalized to their
value at the center of the Dalitz plot. A comparison of the central 
values themselves is shown in 
Table~\ref{tab:centralvalue}.
\begin{figure}
\begin{center}
\includegraphics[height=0.7\textwidth,angle=270]{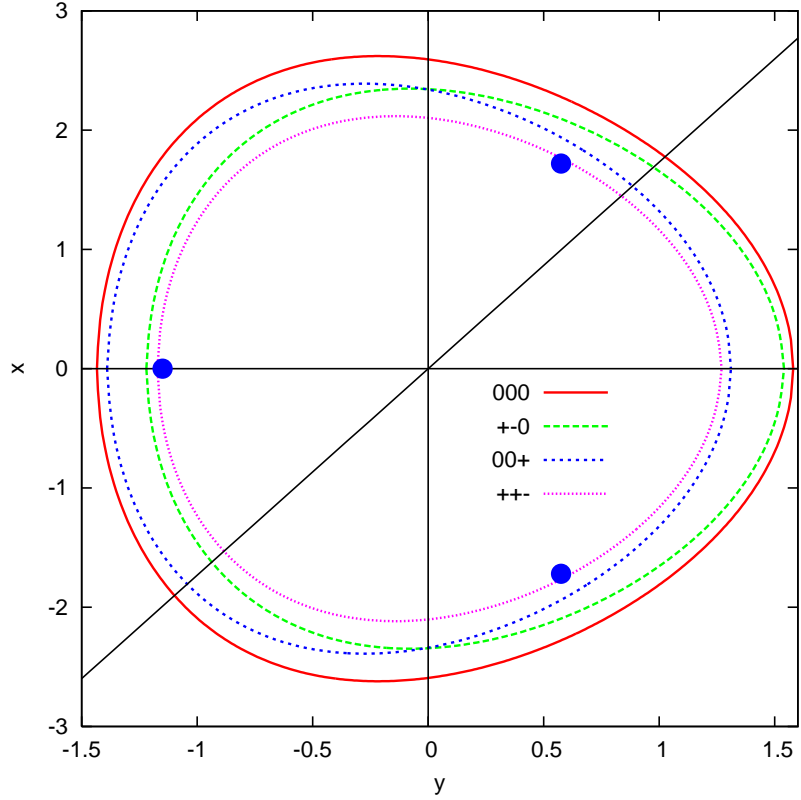}
\end{center}
\caption{The phase space boundaries for the five different decays 
and the three curves 
along which we will compare the squared amplitudes.
The points where two charged pions have low relative velocity
for $K^+\to\pi^+\pi^+\pi^-$ are indicated by the large dots.\label{fig:phase}}
\end{figure}

We also calculate the changes in the Dalitz plot distribution parameters.
These are defined by
\be
\left|\frac{A(s_1,s_2,s_3)}{A(s_0,s_0,s_0)}\right|^2 = 1+gy + hy^2+k x^2\,.
\ee
The isospin breaking corrections to these parameters are given in
Table~\ref{tab:dalitz}. The amplitude for $K_S\to\pi^+\pi^-\pi^0$ 
is parametrized via
\be
A^S_{+-0} = \gamma_S x - \xi_S xy\,.
\ee

\begin{table}
\begin{center}
\begin{tabular}{|c|c|c|c|c|}
\hline
Decay & Quantity & Iso \cite{BDP}& Strong & Full \\
\hline
$K_L \to \pi^0 \pi^0 \pi^0$ & $h$ & $-0.0072$ & $-0.0068$ & $-0.0068$\\
$K_L \to \pi^+ \pi^- \pi^0$ & $g$ & 0.673 & 0.683 & 0.677 \\
                            & $h$ & 0.085 & 0.089 & 0.088 \\
                            & $k$ & 0.0055 & 0.0057 & 0.0057\\
$K_S \to \pi^+ \pi^- \pi^0$ & $\gamma_S$ & $3.4 \cdot 10^{-8}$
 & $3.4 \cdot 10^{-8}$ & $3.5 \cdot 10^{-8}$ \\
 &   $\xi_S$ & $-0.2\cdot 10^{-8}$  & $-0.2\cdot 10^{-8}$& $-0.2\cdot 10^{-8}$\\
$K^+ \to \pi^0 \pi^0 \pi^+$ & $g$ & 0.635 & 0.619 & 0.619 \\
                            & $h$ & 0.074 & 0.071 & 0.071\\
$K^+ \to \pi^+ \pi^+ \pi^-$ & $g$ & $-0.215$ & $-0.211$ & $-0.201$\\
                            & $h$ & 0.012 & 0.012 & 0.008\\
                            & $k$ & $-0.0052$ & $-0.0050$ & $-0.0037$\\
\hline
\end{tabular}
\end{center}
\caption{Comparison of the Dalitz plot distribution parameters
in the isospin conserving case (Iso), including strong and
local electromagnetic (Strong) and full (Full) isospin breaking.
\label{tab:dalitz}}
\end{table}

We now discuss the  results in somewhat more detail. 
In general the results
are of a size as can be expected from this type of isospin breaking.
They are of order a few, up to 11\% in the amplitudes squared
outside the Coulomb
region. The isospin breaking
corrections tend to increase all decay rates somewhat and this will
in a fit
be compensated by small changes in the values of the $\tilde K_i$
compared to the results of \cite{BDP}. The number of significant
digits quoted in Table~\ref{tab:centralvalue} is higher than the expected
precision of our results, but the trend and the general size of the change
compared to the isospin conserving results are stable with respect to
variations in dealing with the eta mass (physical or GMO).

\begin{figure}
\begin{center}
\includegraphics[height=0.8\textwidth,angle=270]{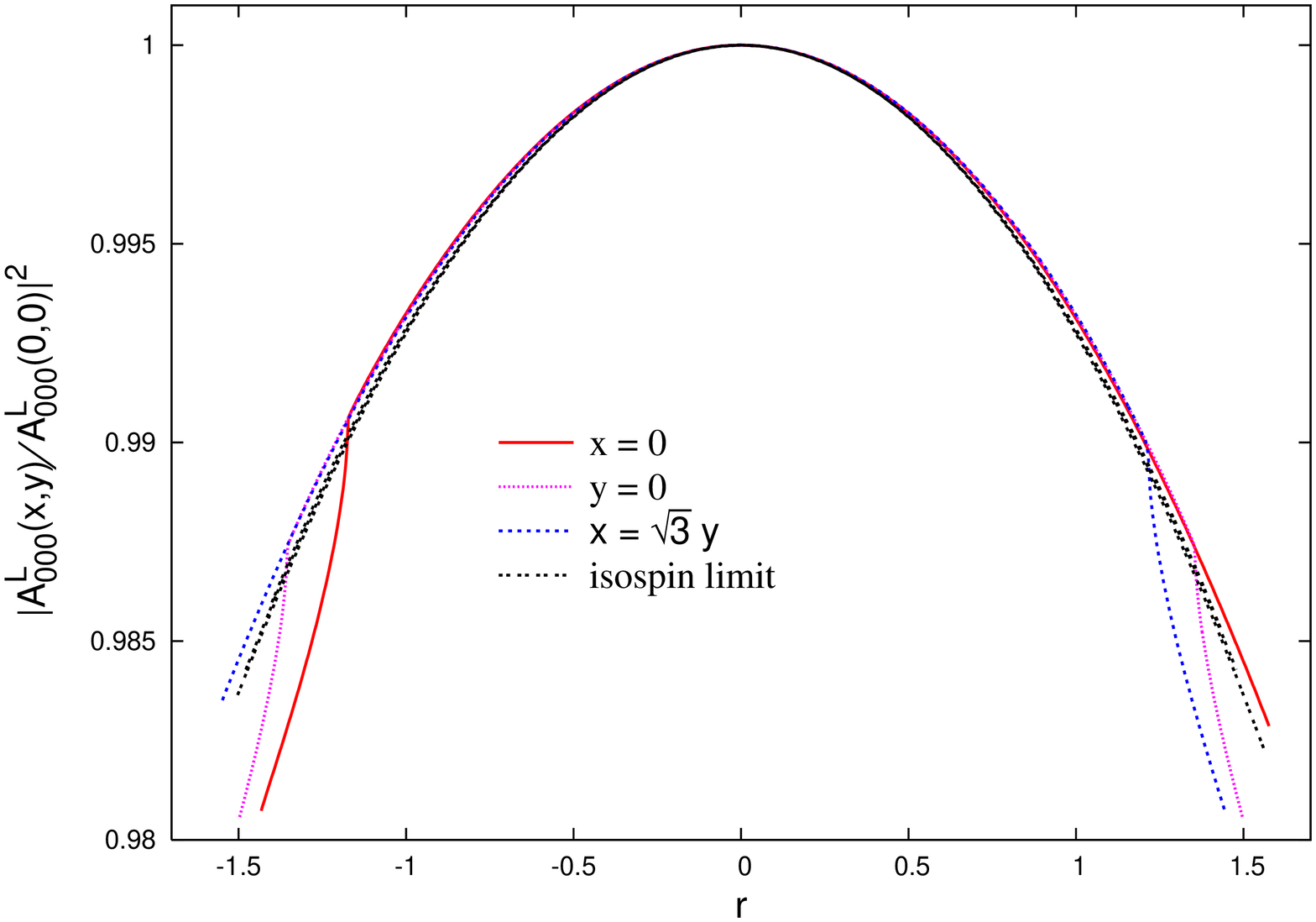}
\end{center}
\caption{Comparison of $K_L\to\pi^0\pi^0\pi^0$ with and without isospin 
breaking.}
\label{fig:compare1}
\end{figure}
For $K_L\to\pi^0\pi^0\pi^0$
the central value of the amplitude squared increases by about 4.5\%. 
In this case we have because of the symmetry of the final state that
$g=0$ and $k=h/3$. The
quadratic slope decreases by about 5\% but the total variation over
the Dalitz plot is small so the total decay rate increases by about 4.5\% as 
well.
This decay is the one which has most variation in the amplitude when
changing how one deals with the eta mass. The extreme case we have found was 
that this effect completely cancelled the change from isospin violation,
but the relative change due to isospin breaking remained similar. 
Note the scale in Fig.~\ref{fig:compare1} when viewing the result.
The changes compared to \cite{BB} are entirely due to the 
photon loop corrections to $F_{\pi^+}$ and $F_{K^+}$.

In Fig.~\ref{fig:compare1} one can also clearly see the 
thresholds induced by the difference between $m^2_{\pi^+}$ and
$m^2_{\pi^0}$ introduced when isospin invariance is broken. These
thresholds correspond to a new process being allowed where two of 
the neutral pions are produced through an intermediate on shell state with one 
positive and one negative pion.  

\begin{figure}
\begin{center}
\includegraphics[height=0.8\textwidth,angle=270]{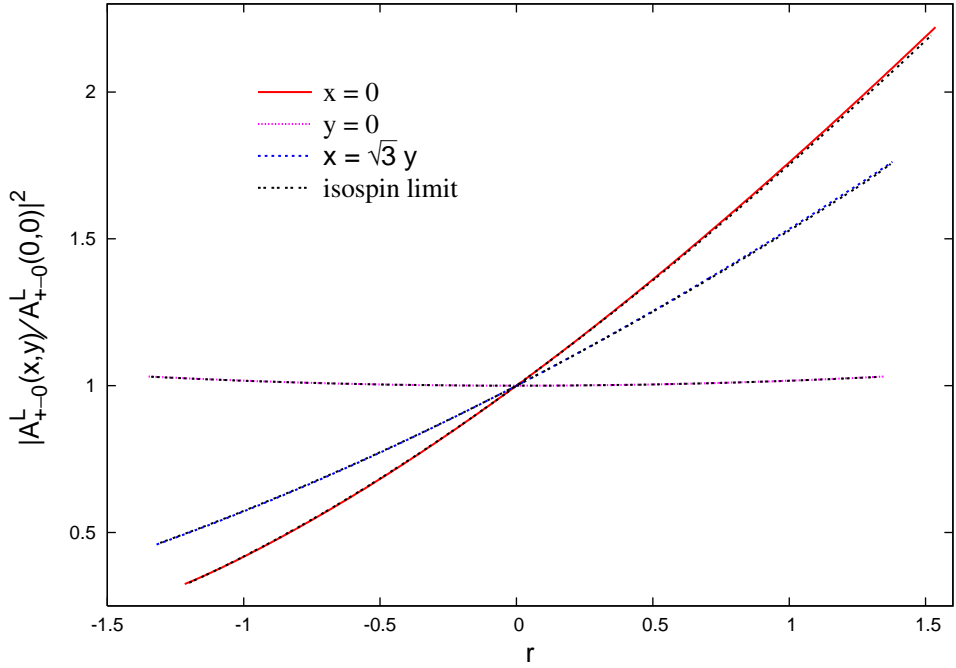}
\end{center}
\caption{Comparison of $K_L\to\pi^+\pi^-\pi^0$ with and without isospin 
breaking.}
\label{fig:compare2}
\end{figure}
The squared amplitude  $K_L\to\pi^+\pi^-\pi^0$ increases by about
5.5\% with very little
variation with the eta mass treatment. The decay rate increases by the same
amount. The changes in the Dalitz plot slopes are rather small as can be
judged from Fig.~\ref{fig:compare2}. The marginal differences compared to $g$
quoted in \cite{BDP} are due to a slightly different fitting procedure to the
amplitudes squared.

\begin{figure}
\begin{center}
\includegraphics[height=0.8\textwidth,angle=270]{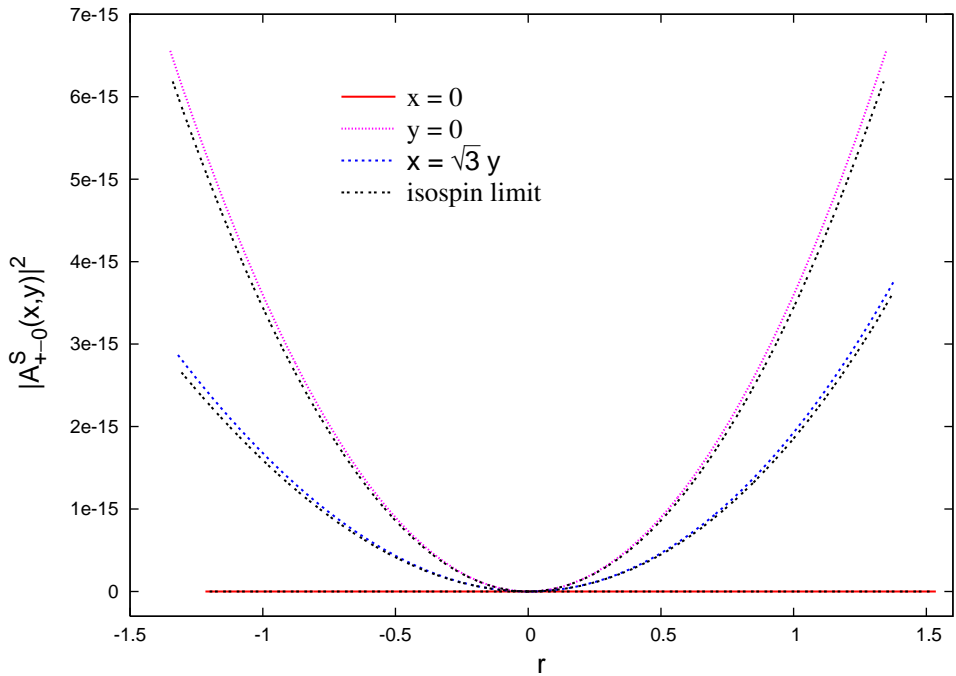}
\end{center}
\caption{Comparison of $K_S\to\pi^+\pi^-\pi^0$ with and without isospin 
breaking.}
\label{fig:compare3}
\end{figure}
For the decay $K_S\to\pi^+\pi^-\pi^0$ the amplitude in the center of the Dalitz
plot vanishes because of the symmetries. The amplitude and the slopes
increase by about 3\% as can be seen in Fig.~\ref{fig:compare3} and the tables.

\begin{figure}
\begin{center}
\includegraphics[height=0.8\textwidth,angle=270]{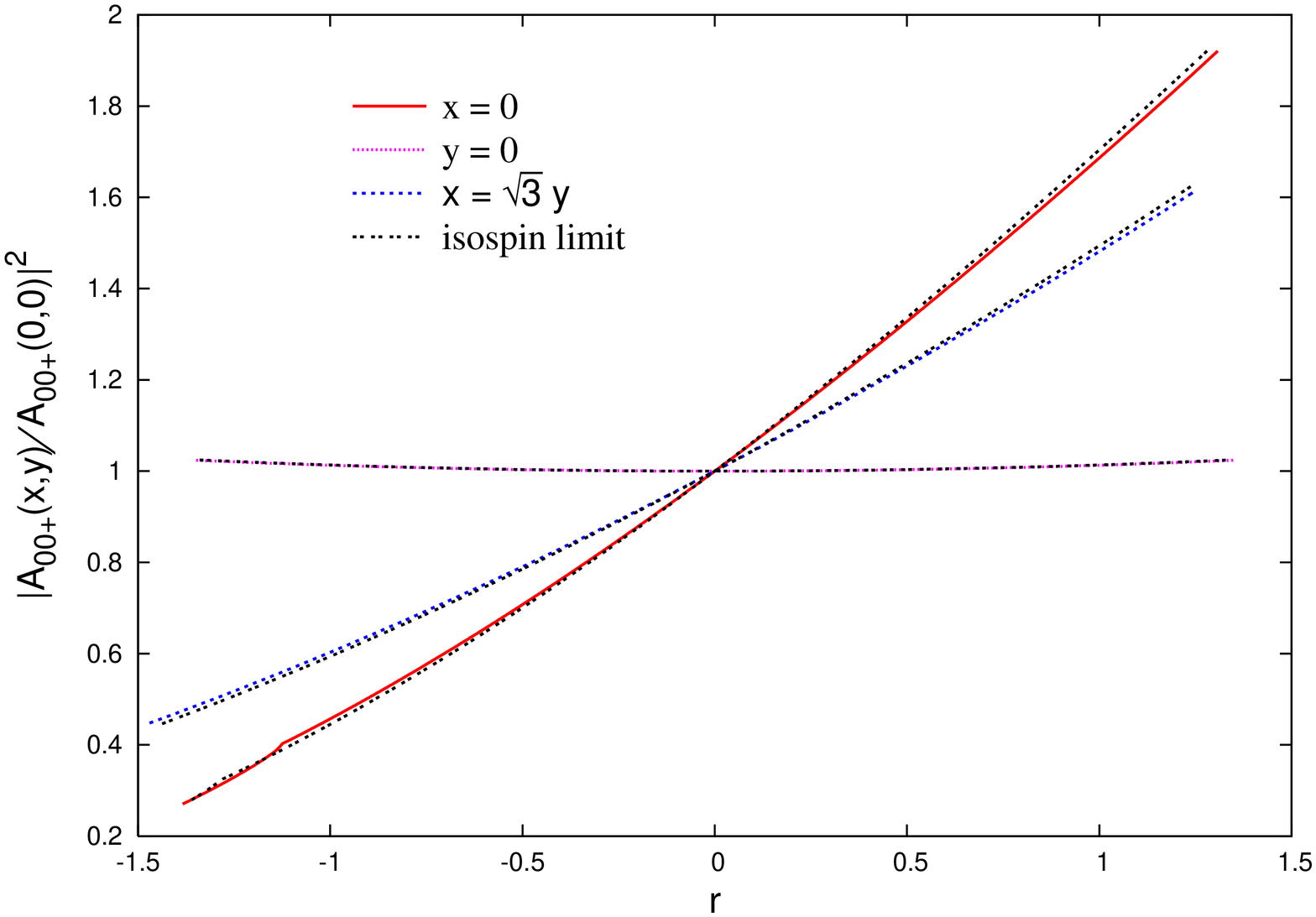}
\end{center}
\caption{Comparison of $K^+\to\pi^0\pi^0\pi^+$ with and without isospin 
breaking.}
\label{fig:compare4}
\end{figure}
The decay $K^+\to\pi^0\pi^0\pi^+$ has a large increase.
The squared amplitude
in the center changes by about 11\%. The linear slopes decrease somewhat
leading to an increase of about 10\% to the total decay rate when compared
with the isospin conserved case. This is shown in Fig.~\ref{fig:compare4}.

The amplitude for $K^+\to\pi^0\pi^0\pi^+$ is also calculated in 
\cite{nehme}. Our
results don't agree with the numerics presented there. We find an
increase in the amplitude while there a decrease is found. 
In \cite{nehme}
a different choice of lowest order was made than here and in \cite{BDP,BB}.
After taking that difference into account, we still disagree significantly
with the numerical results of \cite{nehme}.

\begin{figure}
\begin{center}
\includegraphics[height=0.8\textwidth,angle=270]{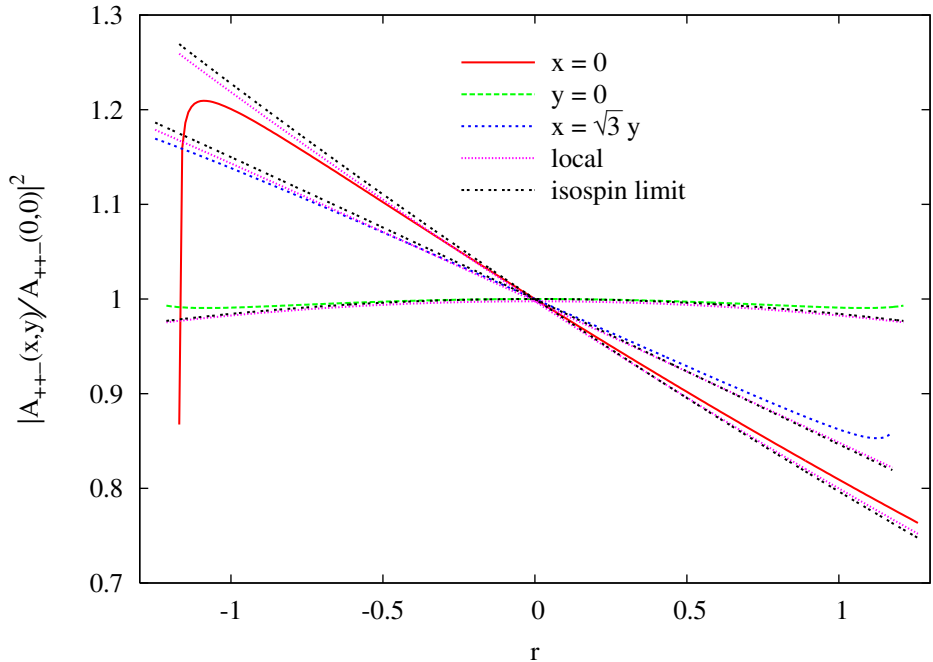}
\end{center}
\caption{Comparison of $K^+\to\pi^+\pi^+\pi^-$ with and without isospin 
breaking.}
\label{fig:compare5}
\end{figure}
The decay $K^+\to\pi^+\pi^+\pi^-$ has a change of about 11\% upwards in
the center of the Dalitz plot.
The slopes decrease somewhat. The decay rate can only be compared when the
Coulomb region is excluded from the comparison but the total change is
also about 11\%.

The conclusions above do not change qualitatively when we give
the $K_i^r$ a value of about $0.001$ and the new isospin breaking  
$\tilde K_i$ a value
relative to $G_8$ and $G_{27}$ of $0.01$. However the changes induced
by these values are numerically significant. They
can be of the order of 10\%, largest for $K_L\to\pi^0\pi^0\pi^0$.

It should be noted that the mentioned changes are with the values of
$\tilde K_i$ determined from the isospin conserving fit in \cite{BDP}. 
A new determination
including isospin breaking effects is planned in an upcoming paper.

\section{Conclusions}
\label{summary}
We have calculated the $K\to3\pi$ amplitudes to
next-to-leading order ($p^4,p^2\,m^2,m^4,p^2\,e^2$) in Chiral Perturbation
Theory. A similar calculation was done in \cite{BDP} in the isospin limit, 
and in \cite{BB} including strong isospin breaking, 
but we have now included full isospin breaking. 
The motivations for this are both because it is
interesting in general to see the importance of isospin breaking
in this process, but also to investigate whether isospin violation will
improve the fit to experimental data. Discrepancies between data and the
quadratic slopes from ChPT were found in \cite{BDP}, and isospin breaking may
be the cause of this.

We have estimated the effects of the isospin breaking by comparing
the squared amplitudes with and without isospin violation. The effect
seems to be at 5-10\% percent level in the amplitudes squared.
To investigate if this removes the discrepancies found in \cite{BDP} 
a new full fit has to be
done, also including the new data 
\cite{istra,kloe} published
after \cite{BDP}. 
This is work in progress
and will be presented in the future paper
Isospin Breaking in $K\to3\pi$ Decays III. 

\section*{Acknowledgments}
The program FORM 3.0 has been used extensively in these calculations
\cite{FORM}. This work is supported in part by the Swedish Research Council
and European Union TMR
network, Contract No. HPRN-CT-2002-00311  (EURIDICE).

\appendix
\renewcommand{\theequation}{\Alph{section}.\arabic{equation}}
\section{The Decay constants $F_{\pi^+}$ and $F_{K^+}$.}
\setcounter{equation}{0}
\label{App:fpi}
We have chosen to normalize our lowest order contribution with 
$F_0^4/(F_{\pi^+}^3F_{K^+})$. $F_{\pi^+}$ and $F_{K^+}$ are the pion and 
kaon decay constants
respectively. Including isospin breaking they are determined from the 
diagrams in Fig.~\ref{fig:decayconst} and the resulting expressions are
\begin{figure}
\begin{center}
\includegraphics[width=0.95\textwidth]{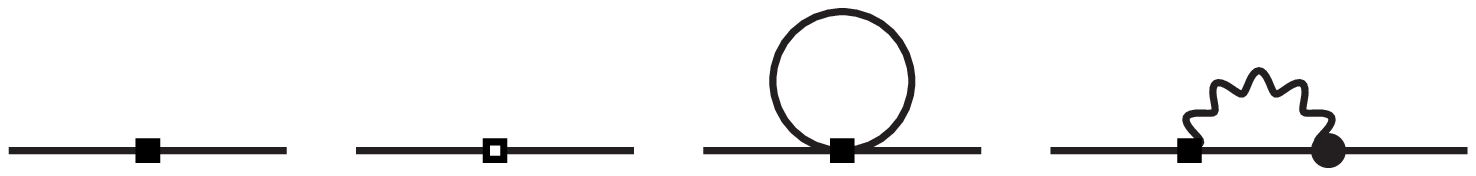}
\end{center}
\caption{Diagrams for the decay constants. An open square 
is a vertex
from ${\cal L}_{W4}$ or ${\cal L}_{W2E2}$, a filled square a vertex
from ${\cal L}_{W2}$ or ${\cal L}_{E2}\, (\Delta S = 1)$ and a filled 
circle a vertex 
from ${\cal L}_{S2}$ or ${\cal L}_{E2}\, (\Delta S = 0)$. A  straight line is 
a pseudoscalar meson and a wiggly line a photon.
\label{fig:decayconst}}
\end{figure}
\ba
\lefteqn{ F_{\pi^+} = F_0\Bigg\{ 1+\frac{1}{F_0^2}\Bigg[ }&&
\nonumber\\&&      
           F_0^2e^2 \, \Big(
           4/3\,K^r_{1}
          + 4/3\,K^r_{2}
          + 10/9\,K^r_{5}
          + 10/9\,K^r_{6}
          + 2\,K^r_{12}
\nonumber\\&&
          + 2\,\frac{\partial}{\partial q^2}\overline{B}(m_\gamma^2,m_\pi^2,m_\pi^2)\,m_\pi^2
          - \overline{B}_1(m_\gamma^2,m_\pi^2,m_\pi^2)
          - 2\,\frac{\partial}{\partial q^2}\overline{B}_1(m_\gamma^2,m_\pi^2,m_\pi^2)\,m_\pi^2
          \Big)
\nonumber\\&&
       + L^r_{4}\, \Big(
           16\,m_\pi^2\,\frac{\sin\epsilon}{\sqrt{3}}
          - 16\,m_K^2\,\frac{\sin\epsilon}{\sqrt{3}}
          + 4\,m_{\pi^0}^2
          + 8\,m_{K^0}^2
          \Big)
\nonumber\\&&
       + L^r_{5}\, \Big(
           4\,m_{\pi^0}^2
          \Big)
       + 1/2\,\overline{A}(m_{\pi^+}^2)
       + 1/2\,\overline{A}(m_{\pi^0}^2)
       + 1/4\,\overline{A}(m_{K^+}^2)
       + 1/4\,\overline{A}(m_{K^0}^2) \Bigg]\Bigg\}
\ea
and
\ba
\lefteqn{ F_{K^+} = F_0\Bigg\{ 1+\frac{1}{F_0^2}\Bigg[}&&
\nonumber\\&&       
F_{0}^2\,e^2 \, \Big(
           4/3\,K^r_{1}
          + 4/3\,K^r_{2}
          + 10/9\,K^r_{5}
          + 10/9\,K^r_{6}
          + 2\,K^r_{12}
\nonumber\\&&
          + 2\,\frac{\partial}{\partial q^2}\overline{B}(m_\gamma^2,m_K^2,m_K^2)\,m_K^2
          - \overline{B}_1(m_\gamma^2,m_K^2,m_K^2)
          - 2\,\frac{\partial}{\partial q^2}\overline{B}_1(m_\gamma^2,m_K^2,m_K^2)\,m_K^2
          \Big)
\nonumber\\&&
       + L^r_{4}\, \Big(
           16\,m_\pi^2\,\frac{\sin\epsilon}{\sqrt{3}}
          - 16\,m_K^2\,\frac{\sin\epsilon}{\sqrt{3}}
          + 4\,m_{\pi^0}^2
          + 8\,m_{K^0}^2
          \Big)
\nonumber\\&&
       + L^r_{5}\, \Big(
           16\,m_\pi^2\,\frac{\sin\epsilon}{\sqrt{3}}
          - 16\,m_K^2\,\frac{\sin\epsilon}{\sqrt{3}}
          + 4\,m_{K^0}^2
          \Big)
\nonumber\\&&
       + 3/4\,\frac{\sin\epsilon}{\sqrt{3}}\, \overline{A}(m_\pi^2)
       + 1/4\, \overline{A}(m_{\pi^+}^2)
       + 1/8\, \overline{A}(m_{\pi^0}^2)
\nonumber\\&&
       + 1/2\,\overline{A}(m_{K^+}^2)
       + 1/4\,\overline{A}(m_{K^0}^2)
       + \overline{A}(m_\eta^2) \, \Big(
           3/8
          - 3/4\,\frac{\sin\epsilon}{\sqrt{3}}
          \Big)\Bigg]\Bigg\}\, .
\ea
These formulas agree with known results \cite{GL2,Fpiem}.

The above formulas are infrared divergent when $m_\gamma\to 0$, and the 
standard way to deal with this is the same as in the amplitudes, i.e.\
adding a bremsstrahlung diagram. We have chosen to just 
add a term including a cut-off
scale for the bremsstrahlung photon,
\be
- |A|_{LO}\,\frac{e^2}{4\pi^2}\,\log \frac{\omega^2_F}
{m_\gamma^2}\,,
\ee 
which cancels the dependence on the photon mass and therefore removes
the divergence. 
The scale $\omega_F$ is set to $10\, \mrm{MeV}$.

\end{document}